\documentclass[runningheads]{llncs}
\usepackage{graphicx}
\usepackage{diagbox}
\usepackage[colorinlistoftodos]{todonotes}

\usepackage{float}
\usepackage{cite}

\begin{document}
\title{Towards a context-dependent\\numerical data quality evaluation framework\\Technical Report}

\titlerunning{Towards a context-dependent\\numerical data quality evaluation framework}
%
%
\author{Milen S. Marev\inst{1}\orcidID{0000-0001-8038-2165} \thanks{This research is funded by EPSRC Doctoral Training Partnership 2016-2017 University of Aberdeen with award number: EP/N509814/1} \and \\
Ernesto Compatangelo\inst{1}\orcidID{0000-0002-8393-3409} \and \\
Wamberto Vasconcelos\inst{1}\orcidID{0000-0001-5090-7581}}
\authorrunning{M. Marev et al.}
%
\institute{Department of Computing Science, University of Aberdeen, Aberdeen, AB24 3UE, United Kingdom  \\
\email{\{r02msm16, e.compatangelo, w.w.vasconcelos\}@abdn.ac.uk}}
\maketitle              
\begin{abstract}

This paper focuses on numeric data, with emphasis on distinct characteristics like varying significance, unstructured format, mass volume and real-time processing. We propose a novel, context-dependent valuation framework specifically devised to assess quality in numeric datasets. Our framework uses eight relevant data quality dimensions, and provide a simple metric to evaluate dataset quality along each dimension. We argue that the proposed set of dimensions and corresponding metrics adequately captures the unique quality antipatterns that are typically associated with numerical data. The introduction of our framework is part of a wider research effort that aims at developing an articulated numerical data quality improvement approach for Oil and Gas exploration and production workflows that is based on artificial intelligence techniques.

\end{abstract}
\section{Introduction -- the quest for numerical data quality}

Numerical data is generated by a variety of sensor and computing sources; they can  either derive from physical measurements (\emph{e.g.}, temperature, speed), or from calculations, or from other algorithmic processes. Whatever the origin of this data,  the quality of a dataset and of each of its individual components can be influenced by a number of different factors (\emph{e.g.}, noise, inaccuracy, imprecision, gaps, or inconsistencies). Some of these factors may critically affect numerical data sets, making the quest for numerical data quality a critical issue in data science, particularly whenever ``big'' datasets are considered. 

Far from being an absolute concept, however, numerical data quality is indeed context-dependent. In fact, in some evaluations and computations a lower quality dataset may still be acceptable. Nevertheless, the very same dataset quality may not be acceptable in other cases, as it could lead to unreliable or meaningless results or worse. For instance, even minor lapses in the expected quality of numerical data used for evaluations and computations in many medical, finance, or industrial scenarios may have catastrophic consequences such as death and/or massive financial losses and/or an environmental disaster. 

To evaluate the quality of a numerical dataset, we need to formally define what ``numerical data quality'' actually means. Although there are various interpretations of this concept~\cite{red2008,deming1991out,dobyns1991quality,juran1990juran}, \emph{we  define numerical data quality as a measure of how much a given dataset adheres to a predefined standard}. Using this broad operative definition, we can introduce metrics to measure numerical data quality in context and decide whether (and if so, how much of) a given  dataset is suitable for a specific computation in a particular context. 
Metrics used to assess the quality of a given dataset with respect to some reference values implicitly define a reference system characterised by a set of non-overlapping \emph{dimensions}. Such a multi-dimensional reference system, together with its associated metrics and with any operational definitions and rules used to pinpoint the context-dependent quality position of a dataset in such reference system, characterises a specific framework for the evaluation of numerical data quality.   

In this paper, we propose a novel framework characterised by eight different numerical data quality dimensions and a simple but effective quality metric associated to each dimension. Both the dimensions and the metrics we use were originally introduced in the context of other frameworks that address data quality for different kinds of numerical and non-numerical data types. These are outlined in the next section. However, our own framework is innovative in terms of both the specific dimensions it uses and the way it uses them. 

The rationale for the introduction of our framework is rooted in our research work aiming at improving the quality of Oil and Gas exploration and production datasets. In this area, the result of computational sequences (denoted as ``workflows'') may often be deemed as unreliable if the quality of the petrophysical  datasets that make up the numerical workflow input falls short of specific high standards. In the circumstances, either data quality improvement techniques (\emph{e.g.}, curation) are used to improve the problematic dataset or, failing this, alternative workflows may have to be pursued which could possibly deliver reliable result without requiring any modification of the input dataset. 
In both cases, identifying and using the most suitable dimensions and metrics for the effective evaluation of data quality in big numeric datasets is of the essence. 


This paper is organised as follows. Section 2 provides an outline of data quality approaches, focusing on the dimensions that are relevant in our numeric context. Section 3 introduces the core features of our numeric data evaluation framework. Finally, section 4 contains an evaluation of the effectiveness of our framework in capturing and addressing relevant  quality issues identified in Oil and Gas exploration and production datasets.



\section{An outline of data quality dimensions} \label{litRev}


Substantial research work in data science has been devoted to data quality; however, most of it focuses on non-numerical data (\emph{e.g.}, text types in database fields). Many of the existing data quality frameworks have introduced their own variant set of data-specific dimensions. In defining our framework, we have selected one or more  dimensions from existing frameworks, adapting them to our numerical data quality focus, and giving rise to a specific data quality reference system of our own. More specifically, we have identified eight data quality dimensions dimensions, namely \emph{accessibility}, \emph{accuracy}, \emph{completeness}, \emph{consistency}, \emph{currency}, \emph{timeliness}, \emph{precision}, and \emph{uniqueness}.

In the more general context of numerical data quality encompassing non-numerical data,  a reference system with more than 20 different dimensions (namely, consistency, volatility, appropriate amount of data, credibility, interpretability, derivation integrity, conciseness, maintainability, applicability, convenience, speed, comprehensiveness, clarity, traceability, security, correctness, objectivity, relevance, reputation, easy of interaction, and interactivity) was proposed in \cite{Batini2009}. Moreover, further dimensions such as believability, concise representation, consistent representation, ease of manipulation, (being) free-of-error, objectivity, and understandability were proposed in \cite{Pipino2002}.

Not all the above dimensions are suitable to address numeric dataset quality. In our framework, we have only retained the eight ones listed above as they are relevant to our numerical universe. However, even considering these numerical data quality dimensions only, there is not a single framework that includes all of the eight ones that we have identified. 
This is because because past data quality frameworks were introduced to deal with data types in (relational) databases, whose records are mostly of non-numerical nature. 
Conversely, several dimensions in our framework (namely, accuracy, consistency, currency, and precision) are of particular relevance in measuring the quality of big numerical datasets, which are often composed of time series. 
This state of affairs is shown in Table~\ref{DimensionTable}.

\begin{table} [htb] 
\resizebox{\linewidth}{!}{%
\centering
\begin{tabular}{ |c|c|c|c|c|c|c|c|c|c|c|c|c|c|c|} 
 \hline
 \centering
\backslashbox[]{Dimension}{Reference}		&\cite{wang1996beyond} &\cite{zmud1978empirical}& \cite{jarke2013}  &\cite{delone1992information}	&\cite{goodhue1995understanding}&\cite{ballou1985modeling}&\cite{wand1996anchoring}	& \cite{Cykana1996DoDGO}& \cite{kovac1997}&\cite{jarke1997} &\cite{redman1998impact}	&\cite{6671378} &\cite{hogan1997accuracy}	&\cite{bailey1983}			\\
\hline
\hline
Accuracy 		& 		+  			  &		+	   				   &		+	 	&		+	 					&		+  	 					&		+	   			  &		   	 				&		+		 		&	 +			 &	+	  			&		+	  				&		+		&							&		+					\\
\hline
Accessibility	& 		+  			  &			   				   &	  	+	 	&		+	 					&		+     					&			   			  &		     				&				 		&				 &		 			&							&				&							&							\\
\hline		
Consistency		& 	 	  			  &			  				   &		+	 	&			 					&		     					&		+	   			  &		     				&		+		 		&				 &	+	  			&		+					&				&							&							\\
\hline
Completeness	& 	 	+ 			  &			   				   &			 	&		+	 					&		      					&		+	   			  &	+	     				&		 +		 		&	+			 &	+	  			& 	+						&				&		+					&		+					\\
\hline
Currency		& 	 	  			  &			   				   &			 	&		+	 					&		 +   					&			   			  &		     				&				 		&				 &		  			&		+					&				&							&		+					\\
\hline
Timeliness 		& 	 	+  			  &			   				   &		+	 	&		+	 					&		      					&		+	   			  &		     				&		 +		 		&	 +			 &	+	  			&							&				&							&		+					\\
\hline
Precision 		& 	 	  			  &			   				   &			 	&								&		      					&		+	   			  &		     				&						&				 &		  			&							&				&							&		+					\\
\hline
Uniqueness 		& 	 	  			  &			  	 			   &			 	&		+	 					&		      					&		   				  &		    				&		 +		 		&				 &		  			&							&				&							&							\\
\hline 
\multicolumn{15}{c}{  } \\
\hline 
\backslashbox[]{Dimension}{Reference}	&\cite{dama_2013}	  &\cite{3540331727}&\cite{Cai2015}		&\cite{Pipino2002}	&\cite{ahituv1980}&\cite{swanson1987}&\cite{olson1982}&\cite{iivari1987}&\cite{de2006analytical}&\cite{Todoran2015}&\cite{loshin_2001}	&\cite{hufford1996}&\cite{even2005value}	&\cite{king1983}					\\
\hline
\hline 
Accuracy 		& 		  			  &			+  	 	&			+ 		&			 		&		+  	      &		   			&		+  	 	  &		+		  &	   +					&	+	  			&	+	 			&		+	&		+		&						\\
\hline
Accessibility	& 		  			  &			+  	  	&	  		+		&		+	 		&		     	  &					&		     	&		+		 &							&	+	  			&		 			&			&				&						\\
\hline	
Consistency		& 	 	  +			  &			+  	  	&			+ 		&			 		&		      	  &					&		     	&				 &							&	+	  			&	+				&			&		+		&						\\
\hline	
Completeness	& 	 	  +			  &			+      	&			 		&		+			&		      	  &					&		     	&				 &		+					&	+	  			& 	+	 			&		+	&		+		&						\\
\hline
Currency		& 	 	  			  &			+     	&			 		&			 		&		      	  &					&		     	&				 &							&		  			&	+	 			&			&		+		&			+			\\
\hline
Timeliness 		& 	 	  +			  &			+	  	&			+ 		&		+	 		&		 +     	  &					&		     	&		 +		 &	 	+					&	+	  			&	+	 			&			&				&			+			\\
\hline
Precision 		& 	 	  			  &			      	&			 		&			 		&		     	  &					&		     	&				 &							&		  			&					&			&				&						\\
\hline
Uniqueness 		& 	 	  +			  &			      	&			 		&					&		      	  & 	 +	   			&		     	&		 		 &						&		  			&		 			&			&				&						\\
\hline
\end{tabular} }
\vspace*{1ex}\caption{Dimension/Reference table (adapted from~\cite{LinLi2012}).}
\label{DimensionTable}
\end{table}

\subsubsection{Accuracy}  
-- This was selected to be part of our quality reference system because numeric data should be as accurate as possible -- namely, its numerical value must be within a specific threshold. The latter is context-dependent, critically depending on the source of numerical data and on its intended usage, thus introducing a strict requirement for correct quality measurements. 
This dimension is particularly critical in assessing the quality of numerical datasets as it is implicitly associated to the number of significant (\emph{i.e.}, \emph{reliable}) decimal digits in a standardised numerical representation. In such representation, each rational number (as all real world numbers end up being once recorded in a digital format) is expressed in the form 
$N = 0.d_1d_2d_3{\ldots}d_m \cdot 10^p$, with $m$ significant digital digits,  where $d_1$ is the (non-zero) most significant one, $d_m$ is the \emph{m--th}, least significant one, and $p$ is an integer exponent.The more the number of significant decimal digits, the more accurate a numerical data item is. 

\subsubsection{Accessibility} -- This
was selected as quite often data must be delivered at the right time, being easily and quickly accessible. Moreover, all corresponding meta-data (data about data) is required to follow the same behavior. 

\subsubsection{Consistency} -- This
was selected to verify that all the data consistently  follows a set of predefined rules (\emph{e.g.}, format, type, structure). 

\subsubsection{Completeness} -- This was selected as any gaps in a numeric dataset may impact on  the result of computational analytics based on averages, best fits, or  regressions, the latter being the basis for devising future trends. 

\subsubsection{Currency} -- This
was selected to ensure that no data resources are wasted and that the datasets selected for processing are the latest. This is where the relevance of this dimension resides, as it specifically designed to deal with dataset ``freshness''. Unfortunately this dimension is not always useful or meaningful in a numerical data context, as quite often observation-derived data refer to some natural phenomenon that cannot be replicated under the very same conditions. This makes each two datasets referring to different instances of the same event, acquired at different times, unique and such that the newer one is certainly not any ``better'' than the older one, unless other factors enter into the equation.

\subsubsection{Timeliness} -- This
was selected to measure if the data is acquired in/evaluated during/adequate for the stated time period. It is a crucial dimension for real-time systems, as data may be required within a specific (typically very small) time frame. In case data is not delivered at the right time, a bottleneck effect may be ensue, as some systems may be forced to remain in an idle state. 

\subsubsection{Precision} -- This
was selected to measure how repeatable numerical data from physical (\emph{e.g.}, sensorial) measurements are. Like currency, precision is not always usable as some measurements are not repeatable for some reason or other. 
It should be noted that precision is conceptually different from accuracy, as the latter measures proximity of the given numeric data element to its ``true'' value.

\subsubsection{Uniqueness} -- This was selected 
to check that an element in the dataset is actually unique and that there are no repeating parameters within the dataset whose element were recorded at the same time. In fact, the presence of repeating parameters in such a dataset may introduce bias.

\section{A novel framework for numerical data quality}

\begin{figure} [H]
   \includegraphics[width=\textwidth]{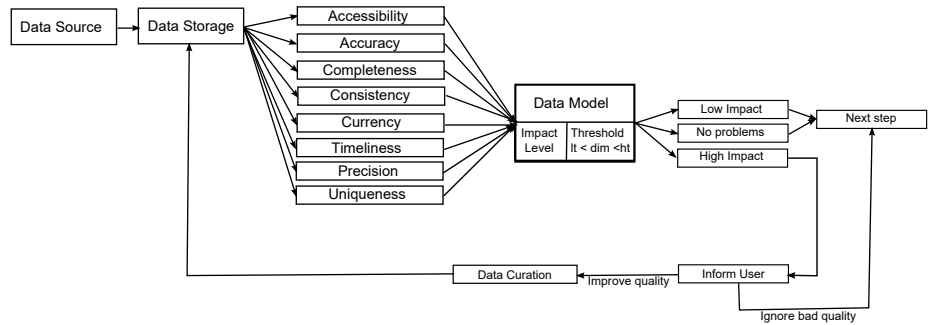} 
    \caption{Numerical Data Evaluation and Data Curation Framework}
    \label{fig:workflow}
\end{figure}

We propose a novel data evaluation framework (NDEF for short) which has been specifically devised to address the assessment of numerical data quality. The introduction of such framework is the first step towards the development of an approach based on artificial intelligence techniques for the improvement of numerical dataset quality. Figure~\ref{fig:workflow} depicts an instantiation of our framework as a set of workflows, each composed of a sequence of nodes (\emph{i.e.}, computational operations). In this paper, we focus on the data evaluation part of the framework (\emph{i.e.}, on the upper side of Figure~\ref{fig:workflow}); the data curation (data improvement) part depicted in the lower side of the figure will be discussed elsewhere. 
Workflows whose nodes represent operations based on numerical calculations are prevalent in science and engineering; for this reason they are often referred to as ``scientific workflows''. The component of the upper part of our workflow-based framework shown in Figure~\ref{fig:workflow} are discussed below.

\subsection{Framework components}

Apart from those components that address each of the eight data quality dimensions, which were discussed in the previous section, the upper part of the framework shown in Figure~\ref{fig:workflow} (namely, the data quality evaluation part) includes other components to manage data quality assessment. These are as follows.

\subsubsection{Workflow input (data source, data storage)} -- Data generated by a variety of heterogeneous information sources (such as sensors or external calculation processes) provide the initial input to the scientific workflow in Figure~\ref{fig:workflow}.  Input data can be permanently recorded in a storage structure such as a database (likely, a NOSQL one in case of big numerical datasets) for later evaluation. 

\subsubsection{Data quality measures, data model and related metrics} --
Depending on the nature of the workflow, different parameters are used in data quality metrics. As shown in  Figure~\ref{fig:workflow}, we use two distinct metrics, namely:
\begin{enumerate}
\item The impact level, which evaluates how important such parameter is for the successful completion of the scientific workflow, and 
\item A threshold with predefined low and high boundaries. 
\end{enumerate}

\subsubsection{Decisional process: progress or improve?} -- The last node in the workflow is a human or computer process to decide whether to progress along that scientific workflow -- either because data is considered to be of good quality in that context or because the considered parameter is deemed to be of low impact in the successful execution of the workflow. Whenever data is deemed of sub-standard quality, the programmatic mechanisms that control workflow execution would inform the user. The latter is prompted to decide whether to improve data quality (and then start the workflow again with improved data) or whether to progress to the next workflow node nevertheless, without performing any quality improvement. In this latter case, their choice would be automatically recorded and reused in further evaluations under the same conditions.

\subsection{Metrics for data quality assessment}

\subsubsection{Accuracy metric} -- This is a measure to calculate the proportion of data that provides a true real-world representation of the phenomenon being measured. To measure dataset accuracy, we use the following formula (adapted from~\cite{even2005value}):
\begin{equation} \label{accEq}
	A_{n,m} = D(X_{n,m}, X^{T} _{n,m})
\end{equation}
where $D$ is the distance between a numerical data item $(X_{n,m}$ and its ``true'' value $X^{T} _{n,m}$ in a collection of $m$ datasets each composed of $n$ values. $X^{T} _{n,m}$, which is  both domain- and case-dependent, is usually defined by an expert. After $D$ has been acquired for all $n$ values in the \emph{m-th} dataset, we can compute the overall accuracy of dataset $m$ using the proportion in formula~\ref{PropEq}, namely  
\begin{equation} \label{PropEq}
	Proportion(m)=\frac{\sum_{i=0}^n A_{n,m}}{n} 
\end{equation}
where $A_{n,m}$ represents the distance of each value from the ``true'' measurement.

\subsubsection{Accessibility metric} -- This measures the proportion of data that is easily, quickly retrievable and available at the right time. 

\subsubsection{Completeness metric} -- This measures the proportion of data that is expected to be recorded against the actual recorded number of values. To measure the completeness of a dataset we use formula \ref{Prop2Eq}, where $m$ is the number of  observed values and $n$ is the  number of expected values.
\begin{equation} \label{Prop2Eq}
	Proportion = \frac{m}{n} 
\end{equation}

\subsubsection{Consistency metric} -- This measures the proportion of data that is consistent and adheres to a predefined standard (which includes format, type, structure, and so on). To measure the consistency of a dataset we first need to define a set of rules that must hold for the dataset (\emph{e.g.}, those rules that make up the standard); they are independent of each other and will have to be considered individually. After the evaluation is complete, we can use  formula \ref{Prop2Eq} to find the proportion of  data elements that adhere to the rules. In this case, $m$ is the number of correct values, while $n$ is the total number of values (taking into account that the dataset includes all the expected values).

\subsubsection{Currency metric} -- This measures the proportion of data that is considered up-to-date and adequate for the selected time period. The base of this metric is defined by an expert who can check if the considered data is still deemed to be ``fresh'' for the purpose of further evaluation. This task can be automated using the specific rule set that describes the data and its validity. To measure the proportion of up-to-date data, we adopt equation \ref{Prop2Eq} where $m$ is the number of current objects, and $n$ number of total observed objects. Here, an object can either be a single data item or a specific subset including several data items.

\subsubsection{Timeliness metric} -- This measures the proportion of data that is delivered to the system at the appropriate time.  It can be measured by comparing the time of the system that is processing the data with the timestamp assigned by the data generating device. In this way, it is possible to detect the delay along the whole communication path between data creation and data reception. Equation \ref{Prop2Eq} is used in this context, where $m$ is the number of timely values and $n$ is total number of received  values.

\subsubsection{Precision metric} -- This measures the proportion of data that is repeatable within the same time period. We use equation \ref{accEq} to calculate precision,  comparing the distance between a number of subsequent samples. In case results are ``distant'' from each other (as defined by an expert threshold value), then precision is lower.

\subsubsection{Uniqueness metric} -- This measures the proportion of identical subsets within each dataset. This situation can occur whenever a dataset is verbatim replicated twice or more in the same columnar structure, as it is possible in NOSQL databases. For instance, these subsets may represent the same parameters within the same time frame. To provide a metric for this dimension we must scan a dataset and identify any subsets  that are being repeated with the same timestamps. All subsets that are not unique will be flagged. Equation \ref{Prop2Eq} is  used to compute this proportion, where $m$ is the sum of all unique subsets and $n$ is the total number of subsets in the data set.

\section{An initial evaluation of our framework effectiveness}
One of our main objectives is to provide a complete and coherent set of numerical data quality dimensions and accompanying metrics, suitable to address any numerical data quality scenario. Hence, we first evaluate all the dimensions introduced in the upper part of Figure~\ref{fig:workflow}. To do so, we consider all the data quality issues that we have identified, namely -- data gaps, added noise, abnormal measurements, and outliers (\emph{i.e.}, abnormal readings far away from other data). 

We need to confirm that all the identified metrics are actually necessary and sufficient to capture any possible \emph{quality antipattern} (each defined as a dataset scenario associated to insufficient data quality in some form or another). Moreover, given a set of metrics based on our NDEF framework dimensions, we would need to perform a quantitative evaluation of our data quality tools, checking whether our dimensions capture all possible antipatterns and effectively measure numerical data quality (or lack thereof) in the assessed datasets.

\subsection{Qualitative evaluation}
There are few unique data quality antipatterns (namely, gaps, noise, and outliers) that can be associated with numerical data. In this section, we try to validate this conclusion and to verify whether the quality dimensions we propose are enough to capture data quality issues and scenarios in their entirety.

\subsubsection{Data gaps} -- 
Their presence means that data is missing -- possibly due to a temporary interruption in a measurement sequence. A gap can either occur once or multiple times within a dataset, with the number of  recorded individual data items being actually smaller than the corresponding number of expected items. Such antipattern is detrimental for  data interpretation as gaps in the information feed would eventually introduce bias in the subsequent processing steps.  As noted in Section \ref{litRev} there are few metrics designed to detect the occurrence of such abnormalities, namely those associated to accessibility and completeness.

\paragraph{Accessibility} -- As already stated, this dimension mainly deals with the availability of the required data at the required time period. In case data is unavailable, it can be deduced that there is a gap in the measurements. Both formulas~\ref{PropEq} and~\ref{Prop2Eq} highlight the data percentage that is available; hence, if formula results indicate a lower value than 100\%, we can conclude that some data is missing. 

\paragraph{Completeness} -- This dimension is quite similar to accessibility, but its metrics have been specifically designed to detect missing readings. When numerical data is  recorded, there is an expectation concerning the structure of recorded readings.  For instance, in a time series, each individual recording must include a data value and  its corresponding timestamp. If the number of recorded values does not match the expected number (\emph{i.e.}, if some timestamps are not associated to any data value or vice versa), then there is a data gap within the dataset. Using the above dimensions, we can make sure that the data gap antipattern is appropriately detected  in the NDEF framework.

\subsubsection{Noise} --
Numerical data can be affected by a variety of different factors that introduce ``noise'' into a dataset. ``Data noise'' can be described as an abnormal variation in  readings/measurements that is detected in a dataset without any obvious reasons for their presence. The most common factors contributing to noise are either of environmental origin (\emph{e.g.}, animals, weather), or of human origin (\emph{e.g.}, wrong interpretations, human errors), or of instrumental origin equipment (\emph{e.g.}, sensor malfunctioning, poor equipment maintenance, wear and tear). Noise can be split into two categories, each with its own characteristics:
\begin{itemize}
\item \textbf {Systematic Errors} -- These affect final results and can impact on value correctness. Such errors can be repeated, as the main source of  the problem is within ``the system''. In the Oil and Gas industry context, for instance, in a number of cases this type of error can be due to heat or pressure. More specifically, when an operator is drilling a new exploration well, the drill bit would heat up because of mechanical friction and thermal conductivity. The drilling pipe would heat up too. The metal the pipe is made of would thus expand due to the higher temperature. This alters the geometry of the pipe, negatively impacting on accuracy and precision of a data reading in a systematic way. Another possibility taken from the same application domain involves a pipe  slightly longer than assumed, with the operator being unaware of the difference. In unconventional wells, where pipe positioning is critical to guarantee hydrcarbon extraction, this may lead to a huge revenue loss, as the operator might miss the correct location of the oil sands. 

Systematic errors can often be easily identified, as  they represent a pattern that repeats itself throughout the dataset in most cases. Systematic abnormalities cause inconsistencies within a dataset and are picked up by a consistency metric. Precision metrics may also be used to detect such problems, notably, if readings are outside a predefined threshold and they do not replicate a patterns identified in previous measurements.
  
  \item \textbf {Random Errors} -- These are among the worst kinds of errors, as they cannot be usually identified and replicated. As the name suggests, they appear at random and without any warning. If not addressed properly, such errors will eventually affect the precision of all data collected. Although hard to detect, this class of errors can be identified using consistency and precision metrics, as they represent value spikes or outliers.
\end{itemize}

\subsubsection{Outdated data}
This data quality anti pattern can be described as data that is not current enough for processing. In such case, this data may not be fit for purpose. For instance, if the address of a customer is not kept up to date in a database, the company will not be able to reach that person, thus wasting company resources. In the numerical domain, outdated data can be considered as data that is not arriving at the processing system by the required time period, because of system bottleneck or communication problems. A currency metric is designed to deal with such errors. Its main purpose is to check how ``fresh'' the data is and whether it is still suitable for further processing. 

Another use of this metric is in detecting whether data is delivered to the processing system by the correct time period. This can be achieved  comparing the data timestamp at its generation point with the system clock. Another metric dealing with this second aspect is timeliness, which is also designed to detect data that is not delivered in a timely manner to the processing system. Both metrics are expected to produce results in the form of a percentage -- illustrating the proportion of data considered out-dated or delayed. Hence, we conclude that these two dimensions are adequate to qualitatively evaluate this kind of scenario.

\subsubsection{Data inconsistency}
This pattern is associated to data that does not adhere to any expected standards like formatting, decimal significance, units, and so on. The metric that is designed to deal with such antipatterns is the consistency one. As previously, explained  this is supposed to detect any data that is not provided in the expected standard, as well as to calculate the proportion of data that is in breach of the expectations. Hence, we conclude that this dimension is adequate to comprehensively evaluate this kind of scenario.

%
%
%

\bibliographystyle{splncs04}
\bibliography{bibliography}

\end{document}